# Advance Reservation of Resources for Task Execution in Grid Environments

Izabela Moise [1,2], Diana Moise [1,2], Florin Pop [1], Valentin Cristea [1]
[1] University "Politehnica" of Bucharest, Romania
[2] IRISA, Campus de Beaulieu, 35042 Rennes Cedex, France
{izabela.moise, diana.moise}@irisa.fr, {valentin, florin.pop}@cs.pub.ro

**Abstract**

*The paper proposes a solution for the Grid scheduling problem, addressing in particular the requirement of high performance an efficient algorithm must fulfill. Advance Reservation engages a distributed, dynamic, fault-tolerant and efficient strategy which reserves resources for future task execution. The paper presents the main features of the strategy, the functioning mechanism the strategy is based on and the methods used for evaluating the algorithm.*

## 1. Introduction

Grid computing represents an emerging technology with important contribution to other contemporary technologies, including application storage provider, storage service provider and peer-to-peer computing. The need for this technology is widespread [6]. There are many domains in today's world which involve the collaboration of individuals, institutes or organizations, which need to access a various collection of heterogeneous resources. This type of large collaborative projects does not imply only the use of high data volumes and data management issues; a main factor in such projects consists of computing resources which go beyond those available to a single person or group. What makes this domain even more challenging is the fact that the resources are often widely distributed: the data sources and storages, the computational resources for analyzing the data and the people interested in the collaborative interpretation of the analysis are situated in different geographical locations, in different *virtual organizations* [1].

In the recent years, many strategies and techniques for scheduling in Grid environments have been developed. Scheduling algorithms used in distributed systems can not be used in Grid environments because of the heterogeneous and dynamic nature of the Grid, characteristics which make the difference between the two types of environments. Scheduling algorithms can be categorized, among others, into static or dynamic, centralized or decentralized [2]. A *static* algorithm assumes that the information about the system resources is available by the time the task is scheduled; on the other hand, if the information is available as the application is executed, then the scheduling is *dynamic*. A dynamic scheduling algorithm is more appropriate to the dynamic nature of the Grid, as the state resources may change in time and the tasks can appear a-periodically.

The recent research on the subject has been directed on both centralized and decentralized strategies for task scheduling. A *centralized* algorithm involves the presence of a single, centralized entity which maintains state information about all the resources of the system. The centralized entity is also responsible for the resource allocation and task scheduling. The main disadvantage this type of algorithm has, is a reduced reliability as it is subject to single points of failure. If the main entity fails, the entire scheduling algorithm fails because it was run by the failed entity. This disadvantage can be overcome if the state information of resources and the scheduling algorithm are managed and maintained locally, on each resource of the system. This type of strategy is a *decentralized* one, having as main advantage, a higher reliability, being less subject to single points of failure. If a resource of the system fails, no critical information is lost; the scheduling algorithm can be run without taking into consideration the failed entity.

An efficient scheduling algorithm can be evaluated through the use of several criteria. An important criterion for evaluating the optimality of a schedule is the property of *load balancing*.

Load balancing represents the process of maintaining balanced workloads across multiple CPUs or systems. This criterion was chosen because a balanced workload across the Grid has a very important influence on the efficiency of a scheduling algorithm. The most important property of a Grid is the vast number of resources available to a user. The load balancing of resources ensures that the resources are used in an efficient way, by exploiting at the maximum all the available resources. Secondly, by using all the available resources, the completion time of the tasks that need to be scheduled reduces considerably.

This paper proposes a decentralized, dynamic and optimal mechanism for task scheduling in Grid environments, called *Advance Reservation*. The remainder of this paper is structured as follows. Section 2 provides a general presentation of Advance Reservation. Section 3 presents the structure of the proposed system and the functioning mechanism of the proposed strategy. In Section 4, we define the performance indicators by which we can assess the efficiency of the proposed algorithm and we give a test example. In Section 5, we assess the efficiency of the algorithm, using the indicators defined in Section 4. In Section 6 we provide an overview of related work. Section 7 summarizes the main ideas of the strategy and outlines directions for future work.

## 2. Advance reservation – a general presentation

In general, Advance Reservation refers to the process of requesting various resources for use at a later time. This method allows Grid resources to be reserved in advance for a designated set of tasks. It uses a decentralized algorithm in which every resource in the environment maintains its state information locally. As opposed to a centralized strategy, Advance Reservation has the advantage of a decentralized mechanism: a higher reliability, as the responsibility for a good functioning is evenly distributed between the resources. The main purpose of Advance Reservation is to obtain an efficient schedule, in terms of resource load and balancing: the optimal schedule would consist of a balanced assignment of tasks to resources, without overloading a resource or overusing it, in comparison to the others. Advance Reservation proposes a mechanism that receives as input, a set of independent tasks and provides as output a schedule, which consists of mappings of each task on the suitable resource of the Grid; the mappings are called *reservations*.

## 3. The functioning mechanism

### 3.1. System anatomy

The system consists of two types of software entities: *the broker* and *the agent*.

*A Broker* represents an entity designated to provide the interface with the user: it receives a set of tasks from the user, builds the final schedule and provides a reply to the user. It also communicates with agents. A broker has an important role (the *decision* – described in Section 3.6) in the scheduling algorithm, but it does not maintain state information about any resource.

*An Agent* maintains state information about the resources it is designated to manage. It monitors its local resources and makes reservations on them. An agent also communicates with brokers, receiving messages and sending replies.

The architectural model of the system is structured as follows: The elementary entity of the system is *the node* or *the resource*. A node consists of one or more central processing units. As this paper proposes an algorithm for task scheduling in Grid environments, the system resources are heterogeneous. A set of nodes forms *a cluster* and a set of clusters forms *a farm*.

### 3.2. Description of a task

A task represents a specific piece of work required to be done as part of a job or application. We consider the following specification tags for a given task:
- taskId represents an unique identifier, by which the task can be referred to;
- startTime and endTime specify the exact moment of time at which the task execution must begin, respectively, the estimated moment of time at which the execution must end; these tags are both expressed in seconds;
- load represents the approximate usage of resources the task requires in order to be executed; the load is specified in percentage;

The specifications for several tasks are contained in XML files, created statically before the running of the algorithm. A user can specify to a broker the name of an XML file, requesting the scheduling of a set of tasks.

### 3.3. Description of a resource

A specification of a resource consists of the following:
- Id represents an unique identifier by which the node can be referred to;
- NodeName specifies the name of the node;
- ClusterName specifies the name of the cluster the node belongs to;
- FarmName specifies the name of the farm the cluster belongs to;

- Parameters specify the characteristics of the processing resource: CPUPower, Memory, CPU_idle;

An agent receives (as a command line argument) the name of an XML file which contains the specifications for several nodes. Each agent maintains state information about the resources it manages, the strategy used by Advance Reservation being a decentralized one.

### 3.4. Communication protocol

We propose a communication protocol that manages the interaction between agents and brokers. Figures 1 and 2 show the *communication model* and, respectively, *the steps of the protocol.*

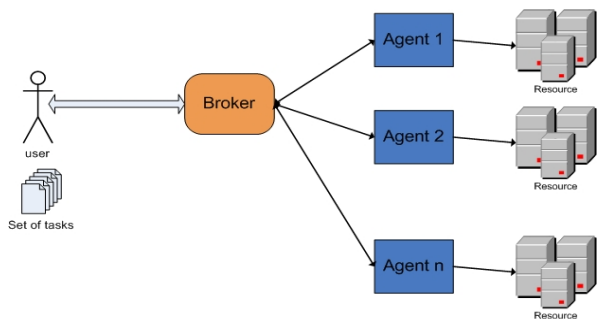

**Figure 1. The communication model**

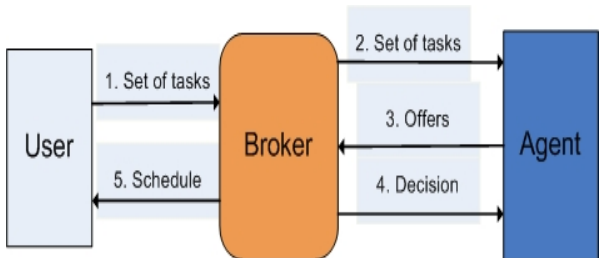

**Figure 2. The steps of the protocol**

The protocol involves the following steps:
*Step 1:*
- the broker receives a set of tasks from the user;

*Step 2:*
- the broker broadcasts the set of tasks to all the agents that are connected in the system, at that time; each agent tries to schedule all the tasks that it received, on its local resources;

*Step 3:*
- each agent sends a reply to the broker, specifying what tasks it was able to map, on which resources and the load each resource would have if a specific task would be executed on it;

*Step 4:*
- the broker receives replies from all agents and builds the final schedule;
- the broker sends the decision to every agent, letting it know what reservations it must make;

*Step 5:*
- the broker provides to the user the final schedule;

### 3.5. Constants and conditions used by advance reservation

In this section, we introduce the definition of three constants and we state two conditions:
- *MAX_LOAD* – defines the maximum value accepted for a resource load; the strategy uses the value 85%, which means a resource can be loaded up to 85% of its full processing capacity (following the model of the *JVM*);
- *MAX_TASKS* – defines the number of tasks which can be run on the same resource at the same time;
- *INFINITE* – defines the maximum value for the superior limit of an interval; the strategy uses the value Long.*MAX_VALUE*;

The two conditions are the following:
1. Each task is defined, among others, by its execution time, given by the interval: [startTime, endTime]. If more than one task have the same execution interval, they can be executed simultaneously on the same processing resource, with the condition that the number of tasks is less than *MAX_TASKS*.
2. A processing resource's load can be increased by scheduling tasks to be executed on it, with the condition that the new load does not become greater than *MAX_LOAD*.

### 3.6. The functions of a broker

The main role of a broker is to obtain, for a given set of tasks, a schedule, consisting of mappings between every task of the set and a resource of the system. The schedule must fulfill optimality criterion: load balancing of the resources in the system. The first thing a broker does when it is launched, is to create a socket on a port (the port number is received as command line argument), on the local machine; the socket will be used for communication with agents.

A broker consists of two threads:

- a thread that receives connection requests from other agents in the system;
- a main thread that accomplishes the following functions:
  1. *providing the user interface*:
     - the broker provides the interface with the user, receiving a set of tasks from the user and providing as output a schedule;
  2. *obtaining the task batch:*
     - the broker waits for the user to enter the name of the XML file which contains the description of one or several tasks;
     - builds a *task batch* which is a vector of tasks; the broker parses the input file and extracts the tasks, building the task batch;
  3. *broadcasting the task batch:* the broker sends the task batch to every agent that is connected, at that specific time;
  4. *waiting for replies;*
  5. *building the final schedule:*
     - the broker builds the final schedule based on the replies from agents; a reply from an agent contains a reservation or a scheduling offer for a given task, also specifying the load of a resource, if the task would be executed on it;
     - the broker keeps track of all the offers received up to a giving moment, in a structure called *finalSched*, which will contain, at the end, the reservations made for the initial set of tasks; when a new offer is received, the broker identifies one of two possible situations; the broker checks to see if it has already received other offers from other agents, for the task the current offer refers to;
       - first situation: the broker has not received any other offers; in this case, the broker adds the current offer to finalSched, in order to compare it to other future possible offers which refer the same task;
       - second situation: the current offer is not the first offer referring to a certain task; in this case, the broker has to make a decision;
  6. *taking decisions:*
     - this step is very important for the optimality of the scheduling; the broker compares the current offer and the existing one; the criterion of comparison is the load of the reserved resource of each offer, the broker choosing the offer which reserves the resource with the lower load value; if the two offers have the same value for the load of their resources, the broker will choose the offer coming from the *"less loaded"* agent; in order to do that, the broker keeps track of how many reservations it has made on every agent that is connected; the "less loaded" agent is the agent which has the lower number of reservations; these two criteria provide the guaranty that, in the end, the broker will obtain a schedule with the property of load balancing;
  7. *broadcasting the decision:*
     - the broker now has the final schedule;
     - it sends, to each agent, the task identifiers of all the tasks that are to be scheduled on the resources of a certain agent; this is the confirmation step, the broker notifies each agent what were the offers it has accepted;
  8. *providing the output;*
  9. *scheduling the remaining tasks:*
     - the broker checks to see if there are tasks which remained unscheduled (tasks for which none of the agents sent offers); if there are, the broker tries to reschedule the tasks, the new task batch will contain the remaining tasks;

### 3.7. The functions of an agent

In order to accomplish their functions, the agents maintain a structure, called *the dynamic table*, which holds information about all the system resources.

**The Dynamic Table**

State information refers to the usage of each resource and it is acquired by monitoring the evolution in time of the resources. The monitoring refers to: the tasks which are scheduled on a specific resource; the interval of time each task requires for its execution; for each interval, the degree of usage of that resource (the load of the resource on that interval).

We describe an interval by the following:
- its inferior and superior limits – start time and end time;
- a vector of tasks which are scheduled to be executed during that interval;
- the usage of the processing resource during this interval;

Each resource has associated a vector of intervals, which are kept in increasing order, after start time. The increasing order is useful in the process of making reservations, when an agents needs to build the specific interval/intervals for a given task.

Therefore, the dynamic table is a mapping between each local resource and a vector of intervals. It contains all the reservations the agents make on the system resources. The dynamic table undergoes changes in time. Initially, the dynamic table contains, for each local resource, the interval [0, *INFINITE*]. Every time an agent makes a reservation, the dynamic table changes, because a new interval must be added to the vector of intervals of a resource or the existing intervals must be changed (the limits, the vector of tasks and the usage must be modified). The dynamic table is kept distributed among all the agents of the system. Each agent maintains the part of the dynamic table which corresponds to its local resources. The table is not managed by a certain entity because it can belong to a different VO than the others. An agent is not aware of other resources, it can monitor only the local resources it was designated to manage. On the other hand, we propose a decentralized strategy, that does not require the presence of a single entity (agent or broker) which maintains state information about all the resources of the system. In Figure 3, we give an example of a dynamic table in case of a system with two resources and seven scheduled tasks:

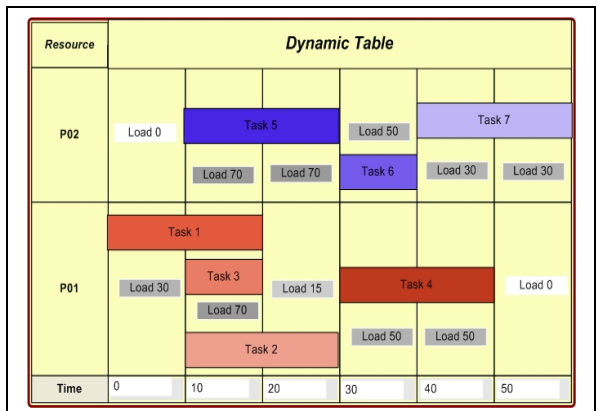

Figure 3. An example of a dynamic table

The first thing an agent does when it is launched, is to create a socket on the IP and port number of a broker (the port number and IP are received as command line arguments); the socket will be used for communication with the broker.

An agent receives the name of an XML file which contains the description of the local resources the agent has to manage. The agent accomplishes the following functions:

1. *obtaining the local configuration;*
2. *initializing the dynamic table:*
   - the agent builds the dynamic table, mapping each of its local resources to a vector of intervals, which initially contains one interval, [0, *INFINITE*], no scheduled tasks and usage 0;
3. *connecting to the broker;*
4. *waiting for a request (a task batch from the broker);*
5. *managing a clone of the dynamic table:*
   - the agent uses a *clone of the dynamic table* for the scheduling algorithm, because only after the broker sends its decision, an agent can "commit" some of the reservations it has proposed; the "committing" step means the agent updates its dynamic table with reservations accepted and confirmed by the broker; the agent must ignore the other reservations it has sent and weren't accepted;
6. *the scheduling algorithm* consists of the following steps:
   - for every received task, the agent tries to find the suitable processing resource; the agent does that by inspecting all its local resources;
   - for each local resource, the agent searches for the suitable interval for the given task;
   - if there are more than one processing resource found, (the task can be scheduled on an interval on each of them), the agent chooses the processing resource that has the *minimum value* for the usage on the suitable interval;
   - the selection criterion for the suitable resource (the minimum load value in case of several available resources), ensures that, in the end, the schedule the agent obtains has the property of load balancing of resources, which will lead to a final schedule with this property;

7. *sending a reply to the broker:*
   - the agent sends a vector of reservations it was able to make;
   - the agent sends offers only for the tasks it was able to reserve resources for, which means the reply does not contain offers for the tasks the agent could not schedule;
8. *waiting for decision;*
   - the agent waits for the final decision from the broker; by the means of this decision, the broker notifies each agent what reservations to commit;
9. *committing the reservations:*
   - when the agent receives the identifiers of the tasks it must schedule, the agent has to modify the dynamic table, according to the confirmed reservations;
10. *monitoring information:*
    - after each set of tasks the agent schedules, it sends information about its local resources to a MonALISA[3] farm; the information consists of: the average load of its local resources, calculated as the arithmetic average of all the loads of that resource, on each of its intervals; the number of tasks the agent schedules on its local resources, after receiving confirmation from the broker;

## 4. Testing and evaluation

In order to assess the proposed algorithm, we introduce the definitions of several performance indicators, which are monitored during each test.

*The evolution of the dynamic table* parameter shows the load of each resource (the usage degree of the resource, expressed in percentage) on each of the execution intervals.

*The load of an agent* indicator is defined by the number of tasks for which the agent made reservations on its local resources. This parameter is monitored by each agent.

*The performance indicator* is defined as follows:

$$\frac{(number\ of\ scheduled\ tasks)}{(total\ number\ of\ tasks)} * 100 .$$

It expresses, in percentage, the amount of tasks the algorithm was able to schedule.

*Communication time* indicator tests the architecture of the system. A broker communicates with agents using Java sockets. The *communication time indicator* is defined as the necessary time for a task batch delivery.

A test scenario is described by the following: the test architecture, the input files, the entities involved in the testing, the performance indicators that the test monitors, the obtained results which are used in Section 5 for the assessment of the algorithm.

Each test requires the presence of at least one broker and one agent. A test example is described as follows:

The test architecture consists of a cluster of 5 different nodes. The cluster name is **Rudolf Cluster** and the nodes are: the main station (called Rudolf), station1, station2, station3 and station4.

The entities involved in the testing are one broker and two agents. The broker is started on the main station, the first agent on a resource which contains station1 and station 2, the second agent on a resource that comprises station3 and station4.

The input file contains 20 task descriptions. The specifications were randomly generated, the tasks have different execution intervals and require different resource load. The obtained results are presented in Section 5.

## 5. Performance evaluation

We use the performance indicators defined in Section 4 to establish if the proposed algorithm fulfills its goal: providing a schedule with the property of *load balancing of resources.*

Some indicators evaluate other parameters of the mechanism, such as the communication time indicator.

### 5.1. The load balancing indicators

This type of indicators (*the evolution of the dynamic table* and *the load of each agent*) have been defined in Section 4. Their role is to establish if the final schedule loads the resources of the system in a balanced manner. The obtained results for the test described in the previous section, are described in the following.

Figure 4 shows the evolution of the dynamic table corresponding to the first agent. The results show that each of the two agents schedules 10 tasks on its local resources, which means we have obtained equal values for the load of an agent indicator.

The performance indicator shows that the algorithm scheduled all the initial tasks, its value being 100%.

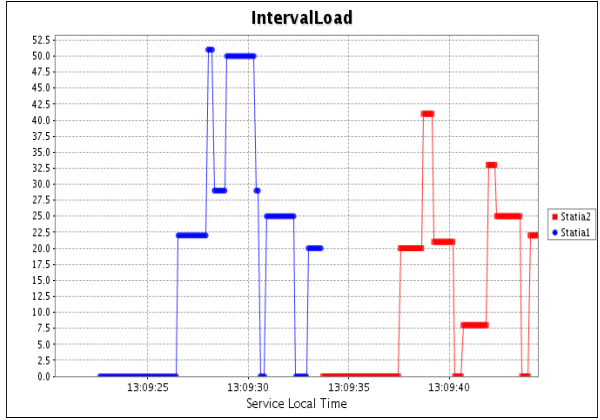

**Figure 4. The evolution of the dynamic table (for the first agent)**

Table 1 shows the experimental results obtained after running tests 1- 4. This table describes the values for indicator *load of an agent*:

**Table 1. Load of each agent**

| Test no. | Agent no. | Load of each agent |
|---|---|---|
| 1 | Agent no.1 | 4 (8) |
|   | Agent no.2 | 4 (8) |
| 2 | Agent no.1 | 10 (20) |
|   | Agent no.2 | 10 (20) |
| 3 | Agent no.1 | 19 (50) |
|   | Agent no.2 | 12 (50) |
|   | Agent no.3 | 19 (50) |
| 4 | Agent no.1 | 36 (100) |
|   | Agent no.2 | 26 (100) |
|   | Agent no.3 | 38 (100) |

The load balancing property of the final schedule is ensured by the algorithm itself. Both the broker and the agent contribute to obtaining this property. In *the decision step* of the algorithm, the broker uses two selection criteria for choosing between two offers from two different agents, which refer the same task:
- *the broker always chooses the offer that less loads the resource;*
- if two offers have the same value for the load tag, *the broker always chooses the offer proposed by the less-loaded agent;*

In case of an agent, when it has to reserve one of its local resources for a given task, it searches for *the less loaded available resource.*

## 5.2. Other performance indicators:

In order to monitor the communication time indicator, we run test no.5, which involves the communication of the file in1.xml, containing 100000 tasks (size of 10MB). The results show that the communication time varies between 5 seconds and 6 seconds.

*The performance indicator* was also defined in Section 4. The tests 1-4 show that the algorithm scheduled all the given tasks, the value for this indicator being 100%.

## 6. Related work

This section presents an overview of the existing scheduling tools that address the problem of advance reservation.

MOAB (Silver) is the most known scheduler that implements a mechanism which reserves resources for a later use [4]. MOAB implements advance reservation for a given set of tasks by mapping each task to the first suitable resource, without taking into account the amount of resources a task requires in order to execute. The Silver toolkit does not allow several tasks to be executed in parallel on the same resource. MOAB uses *Access Control List* to specify a set of rules that manage the access at the reservations.

Calana [5] is an agent-based scheduling system for Grid environments, that uses an auction mechanism for the allocation of resources. The Calana architecture is based on two types of software components: the broker and the agent. The broker has the role of managing the auction process, in order to determine which resource is available whenever a job requires to be scheduled. The agent is run locally on each of the system resources. The agent tries to reserve its resource for the execution of a given task, using this reservation to create a bid, accordingly to the provider's strategy.

## 7. Conclusions

This paper proposes an algorithm for task scheduling in Grid environments, which has as main purpose providing as output, a schedule that is efficient an optimal in terms of load balancing of resources. An important feature of the proposed algorithm is that it uses a distributed mechanisms for monitoring the information state of the system resources, which leads to a higher reliability. The agents are in charge with monitoring the evolution of the system, they do that by maintaining the dynamic table, a structure that is both dynamic and distributed. An important aspect of the strategy is that it allows several tasks to be executed on the same resource, on the same execution interval, fact that increases the number of tasks which are executed simultaneously and decreases the completion time for each task.

There are several possible improvements that can be added to the algorithm:
- the ability to schedule tasks that appear dynamically;
- the synchronization of the access to resources, in case several brokers concur with the same resource;
- the usage of Access Control Lists in order to control access to reservations;

## 8. References

[1] Ian Foster, Carl Kesselman and Steven Tuecke, "The Anatomy of the Grid – Enabling Scalable Virtual Organizations", *International Journal of Supercomputer Applications*, Sage Publications, Inc. Thousand Oaks, CA, USA, 2001, pp. 2-5.

[2] F. Dong and S.G. Akl, "Scheduling Algorithms for Grid Computing: State of the Art and Open Problems", *Technical Report*, School of Computing, Queen's University, Kingston, Ontario, 2006, pp. 7-40.

[3] MonALISA web page - http://monalisa.cern.ch

[4] MOAB web page
http://www.clusterresources.com/pages/products/moab-grid-suite.php

[5] Mathias Dalheimer, Franz-Josef Pfreundt and Peter Mertz, "Agent-based Grid Scheduling with Calana", *Proceeding of the Second Grid Resource Management Workshop*, Springer, Berlin, Germany, 2005.

[6] David W. Walker, "Emerging Distributed Computing Technologies"-
http://users.cs.cf.ac.uk/David.W.Walker/IGDS/GridCourse.htm